\documentclass[conference]{IEEEtran}
\IEEEoverridecommandlockouts
\usepackage{cite}
\usepackage{amsmath,amssymb,amsfonts}
\usepackage{algorithmic}
\usepackage{graphicx}
\usepackage{textcomp}
\usepackage{xcolor}
\usepackage{censor}
\usepackage{multirow}
\usepackage[hyphens]{url}
\usepackage{booktabs}

\def\BibTeX{{\rm B\kern-.05em{\sc i\kern-.025em b}\kern-.08em
    T\kern-.1667em\lower.7ex\hbox{E}\kern-.125emX}}
\begin{document}

\title{WIP: Chat-Debugging: Large Language Model as a Hardware Debugging Assistant\\
\thanks{This work was supported by the National Science Foundation Award EES-2321255.}
}

\author{\IEEEauthorblockN{Andrew Ash and John Hu}
\IEEEauthorblockA{\textit{School of Electrical and Computer Engineering, Oklahoma State University} \\
Stillwater, OK, 78074, USA\\
Emails: \{andrew.ash, john.hu\}@okstate.edu}
}

\maketitle



\begin{abstract}
This work-in-progress research paper explores Chat-Debugging, a novel use case for large language models as an assistant for hardware debugging tasks to improve students' debugging skills. Hardware debugging can be a time-consuming and stressful skill to develop, leading to frustration and other negative emotions. While past work has explored streamlining and automating software-based circuit debugging where digital circuits are dominant, Chat-Debugging aids in physical hardware debugging where circuits may be analog, digital, or mixed-signal. Qualitative data were collected from LLM chat logs and interviews with a fourth-year electrical engineering undergraduate student. Major themes were extracted using a constant comparative analysis. Chat-Debugging incorporates accurate hardware information, properly handles natural language descriptions of circuits, and improves debugging confidence. A successful Chat-Debugging session includes investigating multiple potential root causes proposed by the LLM, the patience and determination to eliminate root causes, and a student who leads the debugging process by assertively correcting the LLM's misunderstandings. This human-computer interaction can improve electrical and computer engineering students' confidence during debugging and improve their debugging skills.

\end{abstract}


\section{Introduction}
Troubleshooting holds different meanings across industries and fields of study. In the semiconductor industry, this takes the form of electronic hardware debugging; this task adds unpredictable time to projects, earning it the nickname ``The Schedule Killer'' \cite{bailey_debug_2021}. Hardware debugging requires some electronics engineers to spend over 40\% of their time troubleshooting circuits \cite{mutschler_debug_2018}. This workforce challenge has roots in education. Electrical and computer engineering students debug circuits during labs and projects throughout their studies, yet, despite its importance, debugging is seldom explicitly taught \cite{crockett_byoe_2025,duwe_defining_2022, dounas-frazer_electronics_2017}. Instead, without scaffolding or guidance, students are left to learn from bugs they encounter. Frustration and confusion from ambiguous debugging expectations may lead students to repeat classes or even change majors \cite{nagvajara_design-for-debug_2007}.

Regardless of the debugging context, if a student lacks the experience to debug a circuit, they need help from someone with more domain knowledge. They might go to classmates, the professor, or post on a forum; regardless, the student spends extra time waiting for help and searching for resources to learn more about the problem. This knowledge gap and the time spent waiting for advice are a natural place to involve a large language model (LLM) in the debugging process. To demonstrate the viability of this educational use case for LLMs, this paper presents the results of an empirical study where a fourth-year undergraduate student used GPT-4o to aid in hardware and software debugging for design projects in their coursework. By comparing and contrasting the familiar use case of software debugging with electrical hardware debugging, the potential of an LLM-assistant to empower students in their debugging process is explored. This paper makes the following contributions:
\begin{itemize}
    \item The first empirical study to use an LLM as a circuit-debugging assistant is presented.
    \item An LLM-assisted debugging protocol is proposed that harnesses the benefits and manages the challenges of this human-computer interaction.
\end{itemize}

The rest of the paper is organized as follows. Section II provides background in electronics debugging and utilizing LLMs in circuit design and debugging. Section III outlines the empirical study and describes the data. Section IV identifies the benefits and challenges of using an LLM as a debugging assistant. Section V concludes the paper.

\section{Background}
\subsection{Challenges of Electronic Circuit Debugging}
Katz and Anderson describe debugging as a specialized form of troubleshooting with four primary steps: understanding the system, testing the system, locating the bug, and fixing the bug \cite{katz_debugging_1987}. While seasoned debuggers reference past experiences \cite{jonassen_learning_2006} and may bypass the process if they have seen the same bug before, experts and novices alike struggle to generate multiple hypotheses to explain unexpected system behavior \cite{alaboudi_using_2020}. Furthermore, lengthy debugging tasks can lead to extensive measurement and testing procedures. Without a thorough documentation process, tests may be unnecessarily repeated, or the test's purpose and the significance of the results may be forgotten \cite{schaafstal_cognitive_2000}. Hypothesis generation and lengthy testing procedures are pain points in the process that could be improved with an LLM-assistant.

The variety of bug location strategies complicates the process. Strategies include discrepancy detection, where deviations from expected behavior are identified; split-half, where the performance of subcircuits is verified; topographic strategies, where signal flow paths are scrutinized; exhaustive strategies, such as completely rebuilding the circuit, where all potential causes are eliminated; and trial and error approaches \cite{crockett_byoe_2025}. While an experienced debugger applies strategies that leverage a system-level view of a circuit, novices may lack the experience to develop complex test plans.

\subsection{Machine Learning for Circuit Design and Debugging}
The design of digital integrated circuits (ICs) before fabrication (pre-silicon) has been streamlined using machine learning (ML). Researchers have automated digital hardware debugging using LLMs to interpret error logs from electronic design automation software, leading to the development of tools for use with hardware description languages \cite{wang_veridebug_2025, li_eda-debugger_2025} and higher-level languages \cite{collini_c2hlsc_2025}. Specialized tools also find and resolve hardware security bugs \cite{ahmad_hardware_2024}. LLM-based tools have even been used to fabricate a CPU IC \cite{wang_chatcpu_2024}.

Adoption of ML tools for analog and mixed-signal circuits has been limited by their complexity, often requiring hand-made designs rather than automated processes. Some tools use ML to interpret circuit schematics in various forms. Xu \textit{et al.} explored image processing to convert a schematic into a netlist (a text-based circuit description) \cite{xu_image2net_2025}. Others have explored how an LLM can identify a circuit's functionality from the netlist \cite{daescu_assessing_2025, pham_genie-asi_2025}. Despite the challenges of analog and mixed-signal design, recent research has begun applying LLMs to pre-silicon circuit design \cite{lai_analogcoder_2025}. Through a combination of written descriptions and graphical analysis, ML methods have even aided in pre-silicon analog debugging \cite{lai_analogcoder-pro_2025}.

The range of research in pre-silicon debugging is indicative of the potential of ML tools for circuit debugging; however, few works explore the verification and debugging that occur after an IC is fabricated. New challenges arise in physical circuits, such as difficulty reproducing errors and the need for slow simulations of large systems to identify the expected behavior \cite{mitra_post-silicon_2010}. One analog hardware debugging algorithm simulates signal perturbations across circuit components to identify potential circuit faults \cite{deyati_atomic_2014}. However, there is no work exploring the potential of LLMs to help debug electronics.

\section{Using an LLM to Enhance Circuit Debugging}
\subsection{Research Questions}
Aspects of debugging that are challenging or overwhelming to a human debugger are well-suited to the strengths of an LLM. To explore this human-AI collaboration, an empirical study was conducted with a fourth-year electrical engineering student given the alias Daniel.\footnote{This human subject research was approved by Oklahoma State University IRB-25-95.} The study explores the following research questions (RQ):

\begin{enumerate}
    \item[] RQ1: What are some benefits that an LLM-assistant provides to a student while debugging circuit hardware?
    \item[] RQ2: What challenges does a student consistently encounter while debugging circuits with an LLM-assistant?
\end{enumerate}

\subsection{Data Sources}
Daniel provided chat logs of conversations with GPT-4o regarding software, hardware/software co-design, and hardware debugging tasks to demonstrate his interaction with an LLM-based debugging assistant in tasks where he self-described as more experienced (hardware) and less experienced (hardware/software co-design and software). He was also interviewed about his debugging processes to gain further understanding of his thought process and perspective on using an LLM for debugging assistance.

\begin{figure}[tb]
  \centering
  \includegraphics[width=\columnwidth]{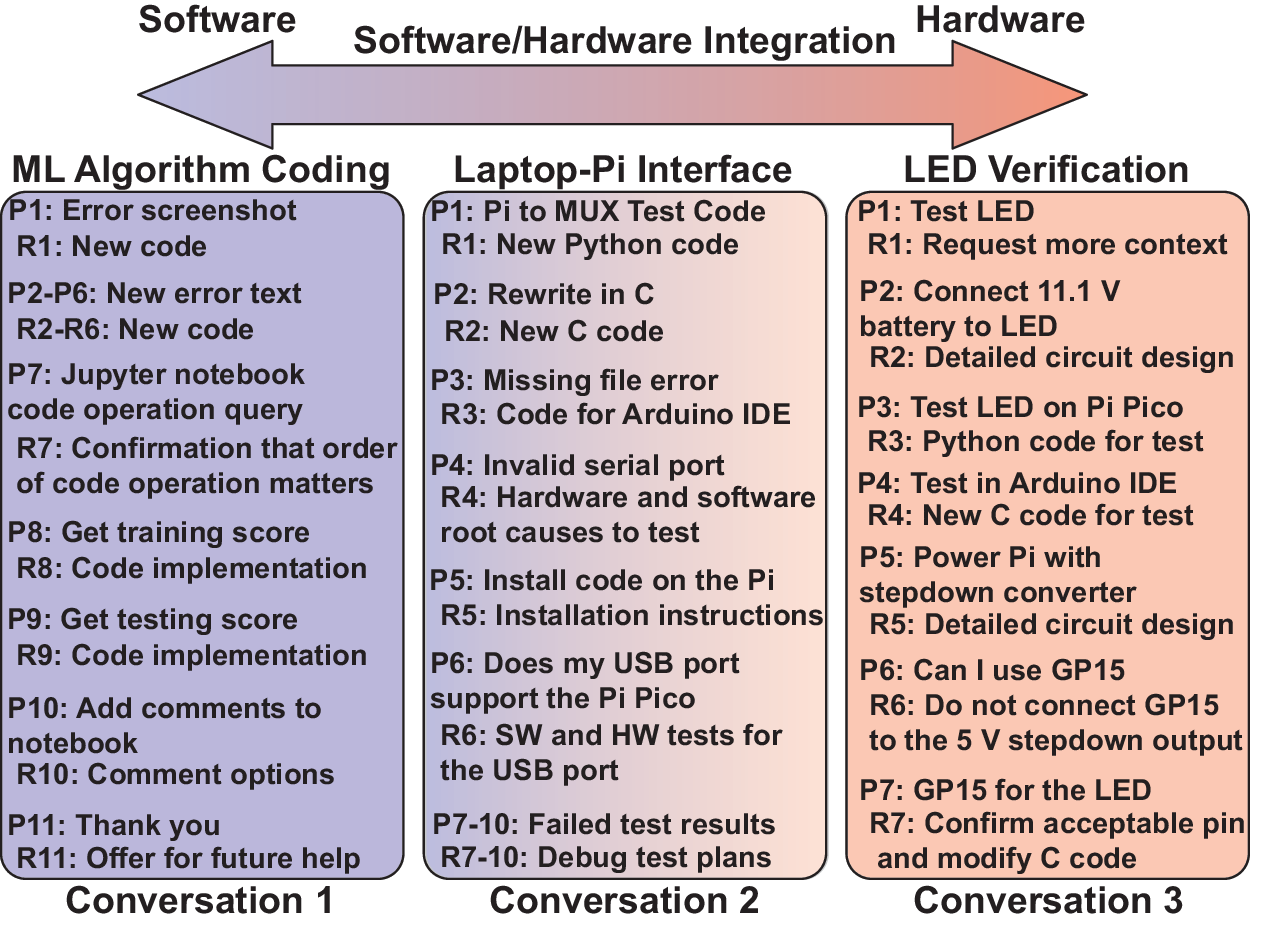}
  \caption{Three chat logs of prompts (P) and responses (R). Conversation 1 is software-focused, Conversation 2 is about software/hardware integration, and Conversation 3 is a hardware debugging task.}
  \label{fig:case_studies}
\end{figure}

The first chat log demonstrates Daniel's interactions with GPT-4o while implementing a machine learning algorithm in a Jupyter Notebook environment. Figure~\ref{fig:case_studies} shows the structure of Conversation 1, a straightforward process with eleven prompts written over a thirty-minute interval. Daniel provided a software error message with no other context, and GPT-4o responded with code that, in most cases, corrected the error on the first attempt. Conversation 1 focused on debugging a series of issues while improving Daniel's domain knowledge of Jupyter Notebook operation and neural network implementations in Python. This conversation does not hold new insights on its own, but acts as a point of comparison for the more hardware-focused debugging conversations.

The second log introduces hardware considerations. Daniel attempted to connect a Raspberry Pi Pico to a multiplexer and verify the connection using C code written for the Arduino Integrated Development Environment. Figure~\ref{fig:case_studies} shows the structure of Conversation 2, where ten prompts were made over a one-hour debugging session. Conversation 2 began with a test script to verify the Pi's connection to a multiplexer, then moved to debugging the hardware connections. Daniel's interactions with the LLM shifted throughout the session. Initially, the LLM assumed the issue could be fixed with software changes; however, after the serial port used for programming the Pi was declared invalid (Conversation 2, Prompt 4), the LLM began considering multiple root causes to isolate the point of failure. Suggestions from the LLM were ordered based on its perceived level of difficulty or the time required for the testing. As software-based root causes diminished, the potential hardware fixes became too complex, forcing Daniel to end the debugging session without resolving the bug. At the time, he did not have extra cables or a second computer available to isolate the hardware error source.

The final chat log is the most hardware-focused. Conversation 3 consisted of seven prompts written over thirty minutes as shown in Figure~\ref{fig:case_studies}. There was a false start as Daniel stated, ``test LED.'' This led GPT-4o to request additional context regarding the hardware setup for the LED test, to which he responded, ``I need to connect an 11.1 V lipo battery to a LED.'' From this point, the LLM explained the potential risks of using a high-power battery on a standard LED and the calculations required to select a current-limiting resistor to maintain safety. Daniel then discussed the implementation of a more complex system: the battery would connect to a step-down converter that supplies 5 V to the Pi that is running a test script to flicker the LED. The LLM identified the correct pins for power and ground connections from the step-down converter, as well as a reasonable general-purpose (GP) pin for the LED. Daniel asked for clarification on whether 15 was an acceptable GP pin selection, which the LLM misinterpreted as a question about acceptable placement for the 5-volt step-down output. Daniel responded by asserting that GP15 would be for the LED. The LLM updated its understanding of the wiring and modified the code with the new pin declaration.

\section{Results}

Two researchers independently reviewed the chat logs and interviews using constant comparative analysis \cite{charmaz_constructing_2006} to identify central themes, then met to discuss and reconcile their findings.

\subsection{RQ1: Benefits of an LLM-Assistant}
Table~\ref{tab:success} summarizes Daniel's successes while using GPT-4o as a debugging assistant with exemplar quotes.

\begin{table*}[tbh]
  \caption{Benefits of Hardware Debugging Using an LLM}
  \label{tab:success}
  \begin{tabular}{cl}
    \toprule
    Success&Exemplar Quotes\\
    \midrule
    1. Accurate Hardware Information&Prompt: ``im trying to verify my pi pico is detecting the mux i have connected to it Im using a PCA9548A''\\
    &GPT: ``Ensure the PCA9548A is properly wired to your Raspberry Pi Pico: [\textit{Connections to power, ground,}\\
    &\ \ \ \ \ \ \ \ \ \ \ \textit{and Pi for the 8 pins needed for the test}]. The default I\textsuperscript{2}C address of the PCA9548A is 0x70...''\\
    &Daniel: ``This pinout is really accurate...these were what it found online as the ideal or just recognizes other\\
    &\ \ \ \ \ \ \ \ \ \ \ people working with Pi Picos''\\
    2. Handles Natural Language Prompts&Prompt: ``I need to connect an 11.1 V lipo battery to a LED''\\
    &Prompt: ``what do i connect the vout from the stepdown to on the pi pico''\\
    3. Boosts Debugging Confidence&Daniel: ``It's super helpful in understanding things, and also, just chugging out mindless tasks.''\\
  \bottomrule
\end{tabular}
\end{table*}

\paragraph{Accurate Hardware Information} Throughout the debugging sessions that Daniel provided logs for, as well as others that he shared clips from during interviews, he consistently found that the LLM provides accurate, detailed descriptions of circuit components. Regardless of whether he asked about an LED, IC, or Raspberry Pi module, the LLM's output provided clear details on the proper connections between components, along with intuitive explanations of each component's intended purpose. On one occasion, Daniel noted that he was debugging the wireless communication between a transmitter and receiver that was failing. The LLM identified key details of the receiver that led Daniel to realize the communication was failing because the component did not meet the targeted specifications; although the circuit could not be modified to fix the issue, through the advice from the LLM, Daniel realized the design would not work as initially planned and needed to be redone.

Beyond accurate wiring descriptions and discussion of device performance parameters, the LLM employed a ``preventative debugging strategy'' for hardware design. It warned about dangerous or common mistakes before discussing the design questions posed to it. For example, it cautioned that the LED test circuit needed to limit the maximum LED current to prevent damage. It identified that a minimum resistor value of 455 $\Omega$ would satisfy this requirement, but suggested using a 470 $\Omega$ resistor, as a standard value Daniel would likely have available. Although it is unclear whether the LLM's understanding of electrical hardware is from training data, searching the internet while crafting a response, or other information sources, its responses regarding hardware components proved robust and helped Daniel throughout his debugging tasks.

\paragraph{Handles Natural Language Prompts} Daniel utilized short prompts such as ``I need to connect an 11.1 V lipo battery to a LED'' and ``it doesnt show up in my microcontrollers.'' Despite brief, informal descriptions to guide its responses, the LLM typically offered relevant information. Only the briefest prompt, ``test LED,'' required more information before a meaningful response was generated. The LLM did not require carefully drafted prompts to respond with guidance. An example from Table~\ref{tab:success} highlights the power of this language processing capability. During his LED test circuit debugging, Daniel prompted ``what do i connect the vout from the stepdown to on the pi pico.'' The structure of the statement could be difficult for a human reader to interpret quickly. There is some ambiguity for a non-technical reader with ``vout'' being shorthand for a voltage output and stepdown being short for a step-down (or buck) converter. Furthermore, this depends on the LLM to understand that Daniel intended to power the device with the step-down converter's output, rather than measuring the output or using it as a control signal. The LLM disambiguated the prompt, interpreted it alongside context from previous prompts, and incorporated knowledge of typical step-down converter use cases to provide detailed help for a bug-free circuit. While natural language processing is not unique to hardware debugging, the complexity of debugging makes this advancement valuable.

\paragraph{Boosts Debugging Confidence} When asked if using an LLM-assistant helped reduce frustration and anxiety related to debugging tasks, Daniel agreed. He cited experiences in which conversations with the LLM helped him understand a topic, trusted it to ``do its own research'' on malfunctioning circuit components and relay the details of the performance issues, and completed other ``mindless'' tasks to free up more time for circuit design work in his senior project. Daniel also found that using an LLM-assistant gave him more confidence in debugging tasks, particularly those where he lacked personal experience. Although this was primarily on the software integration side of the project, he recognized that large electrical systems, where an exhaustive debugging strategy is not feasible, would lead him to turn to the LLM to propose efficient ways to pinpoint the issue.

\subsection{RQ2: Challenges with an LLM-Assistant}
Table~\ref{tab:challenges} summarizes challenges Daniel experienced while using GPT-4o as a debugging assistant with exemplar quotes.

\begin{table*}[tbh]
  \caption{Challenges in Hardware Debugging Using an LLM}
  \label{tab:challenges}
  \begin{tabular}{cl}
    \toprule
    Challenge&Exemplar Quotes\\
    \midrule
    1. Hardware Debugging&Prompt: ``Cannot perform port reset: 1200-bps touch: opening port at 1200bps: Invalid serial port''\\
    Takes Multiple Prompts&GPT: ``1. Incorrect COM Port Selected... 2. Check Serial Connection Settings... 7. Reinstall the RP2040 Drivers''\\
    &\textit{Five prompts with test results and GPT responses follow.}\\    
    2. Requires Consistent&Daniel: ``In the hardware aspect, you really have to take it one step at a time, and it'll be correct.  I can't say, `I have a\\
    Human Feedback&\ \ \ \ \ \ \ \ \ \ Pico and this and this and this and I want to connect all of them.' It'll start putting multiple ones the same pins.''\\
    &Prompt: ``can i use gp15 instead''\\
    &GPT: ``No, GPIO 15 (or any GPIO pin) on the Raspberry Pi Pico is not intended to be used as a power input.''\\
    &Prompt: ``i mean gp15 for the led''\\
    &GPT: ``Ah, I see! Yes, you can use GPIO 15 (GP15) for controlling the LED on your Raspberry Pi Pico...''\\
  \bottomrule
\end{tabular}
\end{table*}

\paragraph{Hardware Debugging Takes Multiple Prompts} Comparing the debugging conversations reveals that hardware debugging with an LLM-assistant requires more patience than software debugging. Software issues were usually resolved in one prompt by providing the error message, incorporating the suggested code, and checking for new errors. Hardware debugging was a more involved process. The LLM offered multiple potential root causes; Daniel then followed the test plans to eliminate root causes and shared the results. These highly interactive conversations continued until the bug was resolved, or time constraints required Daniel to continue the debugging at a later time. For simple hardware debugging tasks, Daniel relied on his embedded systems experience, while the LLM saved time on unfamiliar tasks.

The benefit of this challenge is that it adds mandatory reflection to the hardware debugging process; the LLM outputs cannot be trusted without real-world analysis. Although an LLM suggests incorrect root cause hypotheses, this is similar to the process a student might follow on their own. Potential explanations are explored until they are proved and fixed or disproved and eliminated. \textbf{\textit{Involving an LLM in the process means more potential causes may be considered than a student would propose on their own.}}

\paragraph{Requires Consistent Human Feedback} Daniel identified a key difference in his interactions with the LLM for hardware debugging tasks. He delivered the circuit context in small pieces to avoid overloading the LLM with information that would cause it to misunderstand the complete hardware system. In addition, Daniel had to remind the LLM of complex hardware setups, as its understanding drifted from the real circuit. During long electrical hardware debugging conversations, Daniel found that ``it does start to forget things or modify things that you had already told it.'' However, he also noted that a quick ``don't forget this aspect of it, or don't overlook the fact that I have to do this as well,'' would easily realign the LLM's understanding with reality. \textbf{\textit{If the student assertively maintains the hardware context, consistent human feedback enables an effective circuit debugging assistant.}}

\subsection{An LLM-Assisted Hardware Debugging Protocol}

LLMs are regularly updated and improved, so prompting strategies will change with the models. However, some characteristics of LLM-assisted hardware debugging encountered during this study will likely remain regardless of model changes. Following Katz and Anderson's general troubleshooting model \cite{katz_debugging_1987}, hardware debugging begins with understanding the system. During this step, an LLM-assistant can provide accurate hardware information by summarizing data sheets and describing the expected functionality of various components. A deeper understanding of the system naturally leads to the testing phase, where the LLM can help draft test plans. When the student debugger shares the results of test plans, the LLM can aid in the next step, bug location, by generating potential root causes to explain the incorrect performance identified during testing. Where a human debugging on their own, regardless of experience level, often struggles to generate multiple root causes \cite{alaboudi_using_2020}, including an LLM-assistant leads to a list of potential root causes from training data and internet resources. Generating multiple root cause hypotheses is inherent to circuit debugging; regardless of whether the knowledge comes from personal experience, a classmate, a professor, a forum, or an LLM, root causes must be brainstormed and evaluated. Although an LLM's initial suggestions may be incorrect, the test results can be used to generate more hypotheses until the root cause is found. Once the bug has been located, the student must fix it to return the circuit to expected functionality. While fixing the bug, the LLM may help by reminding the student of relevant hardware context. After this, the LLM can help draft functionality tests to determine if the system is operational. If the tests pass, the student can move forward with improved debugging confidence and new experience from their success.

Hardware debugging takes time, regardless of the resources a student turns to for help. As with any debugging experience, the student should approach the error with patience and endurance. Throughout the process, the student must maintain control of the interaction by validating the LLM's claims to ensure it has an accurate understanding of the hardware. If the student maintains the endurance for debugging and commits to assertively correcting the LLM's misunderstandings, then the challenges present in ChatDebugging can be overcome.

\subsection{Future Work}
This exploratory study used Daniel's self-reported confidence in his debugging results and time spent debugging. The open-ended nature of the data collection led to authentic bugs in Daniel's educational context; however, more participants are needed to improve the validity of the results. We are adding participants by researching Chat-Debugging in a controlled environment where multiple students are given circuits with bugs created by the research team to represent a range of bug categories. The final circuit's performance and time spent debugging are analyzed for quantitative results. This continuation of the research also allows us to explore the suitability of different LLMs to hardware debugging tasks.

\section{Conclusion}
Electronics debugging is ubiquitous in many contexts. While it is inherent to circuit design, it can be more efficient and less uncertain. This work presented an empirical study on the impact of LLM-assisted hardware debugging. The LLM provides and utilizes accurate hardware information, correctly interprets natural language prompts describing electronic hardware, and boosts a student's confidence during debugging. While hardware debugging takes multiple prompts and relies on consistent human feedback to correct misunderstandings of the physical circuit, these challenges help ensure the student actively leads the debugging process, with the LLM acting as an assistant offering advice. These benefits and the positives within the challenges associated with this new human-computer interaction indicate the potential for a more streamlined electronics debugging process to boost students' confidence and improve their debugging skills.

\bibliography{IEEEabrv,references}
\bibliographystyle{IEEEtran}

\end{document}